\documentclass[preprint,preprintnumbers,amsmath,amssymb]{revtex4}
\usepackage{graphicx}

\begin{document}
\title{Gravitoelectromagnetic inflation from a 5D vacuum state:
a new formalism}
\author{$^1$Alfredo Raya\footnote{
E-mail address: raya@ifm.umich.mx},
$^{1}$Jos\'e Edgar Madriz Aguilar\footnote{
E-mail address: edgar@itzel.ifm.umich.mx}and
$^2$Mauricio Bellini\footnote{E-mail address: mbellini@mdp.edu.ar}}

\address{$^1$ Instituto de F\'{\i}sica y Matem\'aticas,
AP: 2-82, (58040) Universidad Michoacana de San Nicol\'as de Hidalgo,
Morelia, Michoac\'an, M\'exico.\\
$^2$ Consejo Nacional de Investigaciones
Cient\'{\i}ficas y T\'ecnicas (CONICET) and Departamento de
F\'{\i}sica, Facultad de Ciencias Exactas y Naturales, Universidad
Nacional de Mar del Plata, Funes 3350, (7600) Mar del Plata,
Argentina.}

\vskip .5cm

\begin{abstract}
We propose a novel
formalism for inflation from a 5D vacuum state which could
explain both, seeds of matter and magnetic fields in the early
universe.
\end{abstract}
\maketitle \vskip .2cm \noindent
\vskip 1cm
\section{Introduction}

It is well known from observation that many  spiral galaxies are endowed with
coherent magnetic fields of $\mu G$ (micro Gauss) strength~\cite{1,2,3,4,5,6}, having
approximately the same energy density
as the cosmic microwave background radiation (CMBR).
For instance, the field strength of our galaxy
is  $B \simeq 3 \times  10^{-6} \  G$, similar
to that detected in high redshift galaxies~\cite{uno} and
damped Lyman alpha clouds~\cite{dos}. There is also evidence for larger-scale
magnetic fields of similar strength within clusters\cite{6'}, which have
been recently reviewed by Giovannini\cite{tres} and by Carilli and
Taylor\cite{cuatro}. These fields can play an important role in various astrophysical
processes, such as the confinement of cosmic rays and the
transfer of angular momentum away from protostellar clouds,
which leads to collapse and formation of stars.
The presence of magnetic fields at even larger
scales has also been claimed\cite{7}.
These fields influence the formation process of large-scale structure\cite{8,9}.
Recently\cite{ocho} the possible existence, strength and structure
of magnetic fields in the intergalactic plane, within the
Local Supercluster, has been scrutinized. The local
supercluster is centered approximately at the VIRGO cluster. A statistically
significant Faraday screening acting on the radio-waves coming from the
most distant sources has been found. This analysis supports the existence
of a regular magnetic field of $0.3 \  \mu G$ in the local supercluster.
More recent discussions of possible observational consequences on
cosmological magnetic fields that include the effects on the CMB anisotropy
were made in\cite{koso}.
Several mechanisms have been proposed to explain the origin of the
seed field. It has been suggested that a primordial field may be
produced during the inflationary period if conformal invariance
is broken\cite{cinco,seis}. In string-inspired models, the coupling
between the electromagnetic field and the dilaton breaks conformal
invariance and may produce the seed field\cite{siete}.

Inflation has nowadays become a standard ingredient for the description
of the early universe. In fact, it solves some of the problems of the
standard big-bang scenario and also makes predictions about CMBR anisotropies which are being
measured with higher and higher precision. The first model of inflation
was proposed by Starobinsky in 1979\cite{star}. A much simpler
inflationary model with a clear motivation was developed by Guth in
the 80's\cite{guth}. However, the universe after inflation in this scenario
becomes very inhomogeneous. These problems were sorted out by Linde
in 1983 with the introduction of chaotic inflation\cite{linde}.
Inflation offers the hope of furnishing a mechanism for kinematically
and dynamically producing the seed of cosmic magnetic fields. It provides
the kinematic means for producing long-wavelength effects in the very
early universe by the amplification of short-wavelength modes of the
inflaton field. This also could have happened with modes of an electromagnetic
field. Since an electromagnetic wave with $\lambda_{phys} \ge H^{-1}_0$ has
the appearance of static $\vec E$ and $\vec B$ fields, very long
wavelength photons ($\lambda_{phys} \gg H^{-1}_0$) could
lead to large-scale magnetic fields.

In this work we shall study a cosmological formalism for inflation
from a 5D vacuum state, where the effective 4D matter, electromagnetic
and vacuum effects are induced geometrically.
The formalism is aimed to explain
both, seeds of matter and magnetic fields in the early
universe.

\section{ 5D Formalism}

We consider the 5D canonical metric\cite{lb}
\begin{equation}\label{eq1}
dS^{2}=\psi^{2}dN^{2}-\psi^{2}e^{2N}dr^{2}-d\psi^{2},
\end{equation}
where $dr^{2}=dx^{2}+dy^{2}+dz^{2}$. In this line
element the coordinates $(N,r)$ are dimensionless and the
fifth one $\psi$ has spatial units. This metric
describes a
5D flat manifold in apparent vacuum $G_{AB}=0$\footnote{In our conventions,
capital Latin indices run from 0 to 4 and greek indices from 0 to 3.} and
satisfies $R^{A}\,_{BCD}=0$, i. e.,
 it's flat.
To describe an electromagnetic field
and  neutral matter on this background, we consider the action
\begin{equation}\label{eq2}
I=\int d^{4}xd\psi\sqrt{\left|\frac{^{(5)}g}{^{(5)}g_0}\right|}
\left[\frac{^{(5)}R}{16\pi {\rm G}} + ^{(5)}{\cal L}(A_{B},A_{C;B})\right]
\end{equation}
for a vector potencial with components
$A_{B}=(A_{\mu},\varphi)$, which
are minimally coupled to gravity.
Here, $^{(5)}R$ is the 5D Ricci scalar, which is zero for the metric
(\ref{eq1}).

We propose a 5D lagrangian density in (\ref{eq2})
\begin{equation}\label{eq3}
^{(5)}{\cal L}(A_{B},A_{B;C})=-\frac{1}{4}Q_{BC}Q^{BC}
\end{equation}
where we define the tensor field $Q_{BC}=F_{BC}+\gamma g_{BC}\left(A^{D}\,_{;D}\right)$,
with $\gamma =\sqrt{\frac{2\lambda}{5}}$ and $F_{BC}=A_{C;B}-A_{B;C}=-F_{CB}$, being $(;)$ the
covariant derivative. The lagrangian density (\ref{eq3}) can also be  expressed as
\begin{equation}\label{ex2}
 ^{(5)}{\cal L}(A_{B},A_{B;C})=-\frac{1}{4}F_{BC}F^{BC}
 -\frac{\lambda}{2}\left(A^{D}\,_{;D}\right)^{2},
 \end{equation}
where the last term is a ``gauge-fixing" term. The
5D-dynamics field equations in a Lagrange formalism
leads to
\begin{equation}\label{eq4}
A^{B}\,_{;D}\,^{;D}-(1-\lambda)A^{C}\,_{;C}\,^{;B}=0.
\end{equation}
Working in the Feynman gauge $(\lambda =1)$, the
equation (\ref{eq4}) yields
\begin{equation}\label{eq6}
\frac{1}{\sqrt{\left|^{(5)}g\right|}}\frac{\partial}{\partial 
x^{C}}\left[\sqrt{\left|^{(5)}g\right|}\,g^{DC}A^{B}\,_{,D}\right]=0,
\end{equation}
where  $A^{B}=(A^{\mu},-\varphi)$. Equation
(\ref{eq6}) is a massless Klein-Gordon-like equation for $A^{B}$
and represents the analogous of the
Maxwell's equations in a 5D manifold in an apparent vacuum.
The commutators for $A^C$ and
$\bar{\Pi}^{B}=\frac{\partial {\cal L}}{\partial (A_{B ,N})}=
F^{BN} - g^{B N}A^{C}\,_{;C}$ are given by
\begin{eqnarray}\label{dif1}
\left[A^{C}(N,\vec{r},\psi),\bar{\Pi}^{B}(N,\vec{r'},\psi ')\right]&=& ig^{CB}
g^{NN} \left|\frac{^{(5)}g_{0}}{^{(5)}g}
\right|\delta^{(3)}\left(\vec{r}-\vec{r'}\right)\delta\left(
\psi - \psi '\right),
\\ \label{dif2}
\left[A_{C}(N,\vec{r},\psi),A_{B}(N,\vec{r'},\psi ')\right]
&=&\left[\bar{\Pi}_{C}(N,\vec{r},\psi),\bar{\Pi}_{B}
(N,\vec{r'},\psi ')\right]=0.
\end{eqnarray}
Here $\bar{\Pi}^N = - g^{NN} \left(A^{C}\,_{;C}\right)$
and $\left|\frac{^{(5)}g_{0}}{^{(5)}g}\right|$ is the inverse of the
normalized volume of the manifold (\ref{eq1}).
From the equation (\ref{dif2}), we obtain
\begin{equation}
\left[ A_C(N,\vec r,\psi), A_{B;N}(N, \vec{r'},\psi')\right] =
- i \  g_{BC} \left|\frac{^{(5)} g_0}{^{(5)} g}\right|
\delta^{(3)}\left(\vec r - \vec{r'}\right) \  \delta\left(
\psi - \psi'\right).
\end{equation}
Using equations (\ref{eq1}) and (\ref{eq6}), the equation of motion for the
electromagnetic 4-vector potential $A^{\mu}$, is given by
\begin{equation}\label{eq9}
\stackrel{\star \star}{A^{\mu}}+3\stackrel{\star}{A^{\mu}}
-e^{-2N}\nabla _{r}^{2}A^{\mu}-\left[4\psi \frac{\partial A^{\mu}}{\partial
\psi}+\psi^{2}\frac{\partial^2
A^{\mu}}{\partial \psi^{2}}\right]=0,
\end{equation}
where the overstar denotes the
derivative with respect to N.
Similarly  for $\varphi$ we have
\begin{equation}\label{eq10}
\stackrel{\star\star}{\varphi} +3\stackrel{\star}{\varphi}-e^{-2N}
\nabla _{r}^{2}\varphi - \left[4\psi\frac{\partial\varphi}{\partial \psi}
+\psi^{2}\frac{\partial ^{2}\varphi}{\partial\psi^{2}}\right]=0.
\end{equation}

Furthermore, using (\ref{dif1}) the commutator between $\varphi$ and $\stackrel{\star}{\varphi}$
becomes
\begin{equation}\label{eq12}
\left[\varphi (N,\vec{r},\psi),\stackrel{\star}{\varphi}(N,\vec{r'},
\psi ')\right]
=i\left|\frac{^{(5)}g_{0}}{^{(5)}g}\right|
\delta^{(3)}\left(\vec{r}-\vec{r'}\right)\delta\left(\psi-\psi '\right).
\end{equation}
which is the same expression that whole
obtained in \cite{MB1}.

\subsection{The  4D Electromagnetic Field embedded in 5D. }

Transforming $A^\mu$ according to
$A^{\mu}(N,\vec{r},\psi)=e^{-3N/2}\left(\frac{\psi _{0}}{\psi}\right)^{2}
\bar{A}^{\mu}(N,\vec{r},\psi)$, and from the equation (\ref{eq9}), we have
\begin{equation}\label{eq13}
\stackrel{\star\star}{\bar{A}^{\mu}}-
e^{-2N}\nabla _{r}^{2}\bar{A}^{\mu}
-\psi^{2}\frac{\partial ^{2}\bar{A}^{\mu}}{\partial \psi^{2}}
-\frac{1}{4} \bar A^{\mu}=0,
\end{equation}
so that the commutator between $\stackrel{\star}{\bar{A}^{\mu}}$
and $\bar{A}^{\nu}$ becomes
\begin{equation}\label{eq14}
\left[\stackrel{\star}{\bar{A}^{\mu}}(N,\vec{r},\psi),
\bar{A}^{\nu}(N,\vec{r'},\psi ')\right]
=ig^{\mu\nu}\delta^{(3)}\left(\vec{r}-\vec{r'}\right)\delta
\left(\psi - \psi '\right).
\end{equation}
The redefined electromagnetic field $\bar{A}^{\mu}$ can be
expressed in terms of a Fourier expansion
\begin{small}
\begin{equation}\label{eq15}
\bar{A}^{\mu}(N,\vec{r},\psi)=\frac{1}{(2\pi)^{3/2}}
\int d^{3}k_{r}\int dk_{\psi}\sum _{\alpha =0}^{3}\epsilon _{(\alpha)}^{\mu}
\left[a_{k_{r}k_{\psi}}^{(\alpha)}e^{
i\vec{k}_{r}\cdot\vec{r}}\zeta _{k_{r}k_{\psi}}(N,\psi)
+a_{k_{r}k_{\psi}}^{(\alpha)\,
\dagger}e^{-i\vec{k}_{r}\cdot\vec{r}}\zeta _{k_{r}k_{\psi}}^{ *}(N,\psi)\right]
\end{equation}
\end{small}
where the creation and annihilation operators
($a_{k_{r}k_{\psi}}^{(\alpha)\dagger}$, $a_{k_{r}k_{\psi}}^{(\alpha)}$)
comply
\begin{eqnarray}\label{eq16}
\left[a_{k_{r}k_{\psi}}^{(\alpha)},a_{k'_{r}k'_{\psi}}^{(\alpha ')\,
\dagger}\right] &=& - g^{\alpha \alpha '}\delta ^{(3)}\left(\vec{k}_{r}
-\vec{k'}_{r}\right)\delta\left(k_{\psi}
-k'_{\psi}\right),\\
\label{eq17}
\left[a_{k_{r}k_{\psi}}^{(\alpha)},a_{k' _{r}k' _{\psi}}^{(\alpha ')}
\right]
&=&\left[a_{k_{r}k_{\psi}}^{(\alpha)\,\dagger},
a_{k' _{r}k' _{\psi}}^{(\alpha ')\,\dagger}\right]=0,
\end{eqnarray}
and the four polarisation 4-vectors $\epsilon ^{\mu}_{(\alpha)}$ satisfy $\epsilon _{(\alpha)}\cdot\epsilon _{(\alpha 
')}=g_{\alpha\alpha '}$.
Using the expansion (\ref{eq15}), into the equation (\ref{eq13}), we find
\begin{equation}\label{eq19}
\frac{\partial^2}{\partial N^2}
\zeta _{k_{r}k_{\psi}}+\left[e^{-2N}k_{r}^{2}-\frac{1}{4}\right]
\zeta _{k_{r}k_{\psi}}-\psi ^{2}\frac{\partial^2}{\partial \psi^{2}}
\zeta _{k_{r}k_{\psi}}=0,
\end{equation}
which is the dynamical equation for the modes
$\zeta _{k_{r}k_{\psi}}$ of
$\bar{A}^{\mu}$.
We propose that $\zeta _{k_{r}k_{\psi}}$ can be
decomposed as $\zeta _{k_{r}k_{\psi}}(N,\psi)
=\zeta _{(1)}(N)\zeta _{(2)}(\psi)$, where for
simplicity, we have suppressed the underscripts $k_{r}k_{\psi}$ in the notation.
Thus, equation (\ref{eq19}) can be equivalently expressed by the system of equations
\begin{eqnarray}
\psi^2\frac{d^{2}}{d\psi^{2}}\zeta _{(2)}
-\kappa^{2}\zeta _{(2)}&=&0, \label{eq20} \\
\frac{d^{2}}{dN^{2}}\zeta _{(1)}+\left[k_{r}^{2}e^{-2N}-\left(
\kappa^{2}+\frac{1}{4}\right)\right]\zeta _{(1)}&=&0,
\label{eq21}
\end{eqnarray}
where $\kappa$ is a dimensionless separation constant given by
$\kappa^2=\psi^2k^2_\psi$, being
$k_{\psi}$ the wavenumber corresponding to the fifth coordinate.
The general solution for the system (\ref{eq20},\ref{eq21}) is given by
\begin{equation}\label{eq22}
\zeta _{k_r k_{\psi}}(N,\psi)
=C_{1}{\cal H}_{\nu _{1}}^{(1)}[x(N)]+C_{2}{\cal H}_{\nu _{1}}^{(2)}[x(N)],
\end{equation}
where $\nu _{1}={\sqrt{4k^2_{\psi}\psi^2 +1}\over 2}$  is
a dimensionless constant and $x(N)=k_{r}e^{-N}$. In this equation
${\cal H}_{\nu _1}^{(1)}$ and ${\cal H}_{\nu _{1}}^{(2)}$ are the
first and second kind Hankel functions. The normalization condition for
$\zeta _{k_{r}k_{\psi}}(N,\psi)$ becomes
\begin{equation}\label{renor1}
\zeta _{k_{r}k_{\psi}}\stackrel{\star}{\zeta }_{k_{r}k_{\psi}}^{\,*}
-\stackrel{\star}{\zeta}_{k_{r}k_{\psi}}\zeta _{k_{r}k_{\psi}}^{*}=i.
\end{equation}
Therefore, considering the Bunch-Davies vacuum, $C_{1}=0$ and
$C_2=i \sqrt{\pi}/2$, we obtain
\begin{equation}\label{solucion}
\zeta _{k_{r}k_{\psi}}(N,\psi)=i\frac{\sqrt{\pi}}{2}{\cal H}_{\nu _1}^{(2)}[x(N)],
\end{equation}
which gives
the normalized modes corresponding to the
electromagnetic field embedded in a 5D aparent vacuum.

\section{Effective 4D dynamics.}
Considering the coordinate transformations
\begin{equation}\label{eq23}
t=\psi _{0}N,\quad R=r\psi _{0},\quad \psi=\psi,
\end{equation}
equation (\ref{eq1}) takes the form
\begin{equation}\label{eq24}
dS^{2}=\left(\frac{\psi}{\psi _0}\right)^{2}
\left[dt^{2}-e^{2t/\psi _0}dR^{2}\right]-d\psi ^{2} ,
\end{equation}
which is the Ponce Leon metric that describes a 3D spatially
flat, isotropic and homogeneous
extension to 5D of a Friedmann Robertson Walker (FRW)
line element in a de Sitter expansion.
Here $t$ is the cosmic time and $dR^{2}=dX^{2}+dY^{2}+dZ^{2}$.
Now, we can take the foliation $\psi=\psi _0$ in
(\ref{eq24}), such that we obtain the effective 4D metric
\begin{equation}\label{eq25}
dS^{2}\rightarrow ds^{2}=dt^{2}-e^{2H_{0}t}dR^{2},
\end{equation}
which describes a 3D spatially flat, isotropic and homogeneous
de Sitter expanding Universe
with a
constant Hubble parameter $H_0=1/\psi _0$ and a 4D scalar curvature $^{(4)}{\cal R}=12H^{2}_0$.

Equation (\ref{eq9}) with the transformation (\ref{eq23})
and the foliation $\psi=\psi _{0}=H_{0}^{-1}$, provides the
effective equation of motion for $A^{\mu}(t, \vec R, \psi=\psi_0) \equiv
A^{\mu}(t, \vec R)$
\begin{equation}\label{eq26}
\ddot{A}^{\mu}+3H_{0}\dot{A}^{\mu}-
e^{-2H_{0}t}\nabla _{R}^{2}A^{\mu}-H_{0}^{2}\left.
\left[4\psi\frac{\partial A^{\mu}}{\partial \psi}+\psi^{2}\frac{\partial ^{2}A^{\mu}}{\partial \psi ^{2}}
\right]\right|_{\psi=H_{0}^{-1}}=0,
\end{equation}
where $A^{\mu}$ is the effective 4D electromagnetic field
induced onto the hypersurface $\psi=H_{0}^{-1}$.
Note that the last term between brackets acts as an induced electromagnetic
potential derived with respect $A^{\mu}$. This term is the analogous
to $V'(\varphi)$ in the case of
an inflationary scalar field as used in \cite{madbe}, and in our
case the dynamics of the component $A^{4} \equiv -\varphi$ is described by
\begin{equation}\label{eq26'}
\ddot{\varphi}+3H_{0}\dot{\varphi}-
e^{-2H_{0}t}\nabla _{R}^{2}\varphi-H_{0}^{2}\left.
\left[4\psi\frac{\partial
\varphi}{\partial \psi}+\psi^{2}\frac{\partial ^{2}\varphi}{\partial
\psi ^{2}}
\right]\right|_{\psi=H_{0}^{-1}}=0.
\end{equation}

On the other hand, transforming $A^\mu$ as
$A^{\mu}(t,\vec{R})=e^{-\frac{3}{2}H_{0}t}{\cal A}^{\mu}(t,\vec{R})$,
the equation (\ref{eq26}) takes the form
\begin{equation}\label{eq28}
\ddot{{\cal A}}^{\mu}-e^{-2H_{0}t}
\nabla _{R}^{2}{\cal A}^{\mu}-\left(\frac{9}{4}H_{0}^{2}
+\alpha\right){\cal A}^{\mu}=0,
\end{equation}
where $\alpha=k_{\psi _{0}}^{2}-2H_{0}^{2}$ is
a constant parameter. Expressing ${\cal A}^{\mu}(t,\vec{R})$ as a Fourier expansion
\begin{equation}\label{eq29}
{\cal A}^{\mu}(t,\vec{R})=\frac{1}{(2\pi)^{3/2}}
\int d^{3}k_{R}\int d_{k_{\psi}}\sum _{\gamma =0}^{3}\epsilon _{(\gamma)}^{\mu}\left[a_{k_{R}}^{\gamma}e^{i\vec{k}_{R}
\cdot\vec{R}}Q_{k_{R}}(t)+cc\right]\delta(k_{\psi}-k_{\psi _{0}}),
\end{equation}
where $k_{\psi _0}$ is a constant. The equation of motion for
the effective 4D electromagnetic modes $Q_{k_{R}}(t)$, becomes
\begin{equation}\label{eq30}
\ddot{Q}_{k_{R}}+\left[k_{R}^{2}e^{-2H_{0}t}
-\left(\frac{9}{4}H_{0}^{2}+\alpha\right)\right]Q_{k_{R}}=0,
\end{equation}
whose general solution is
\begin{equation}\label{eq31}
Q_{k_{R}}(t)=F_{1}{\cal H}_{\nu}^{(1)}[y(t)]+F_{2}{\cal H}_{\nu}^{(2)}[y(t)],
\end{equation}
where $y(t)=\frac{k_{R}}{H_{0}}e^{-H_{0}t}$
and $\nu=\frac{1}{2H_{0}}\sqrt{9H_{0}^{2}+\alpha}$.

The corresponding normalization condition for the modes $Q_{k_{R}}(t)$
becomes
\begin{equation}\label{eq32}
Q_{k_{R}}\dot{Q}_{k_{R}}^{*}-\dot{Q}_{k_{R}}Q_{k_{R}}^{*}=i.
\end{equation}
Note that $H_0$ remains constant in a de
Sitter expansion.
Therefore, taking into account the Bunch-Davies vacuum condition, we
consider $F_{1}=0$ and $F_2=i \sqrt{{\pi\over 4 H_0}}$.
Hence the normalized solution of (\ref{eq30}) is
\begin{equation}\label{eq33}
Q_{k_{R}}(t)=i\sqrt{\frac{\pi}{4H_0}}{\cal H}_{\nu}^{(2)}[y(t)].
\end{equation}
which describes the normalized effective 4D-modes corresponding to the effective 4D electromagnetic field $A^{\mu}$.

\subsection{Classicality conditions of $A_{\mu}$}

It is very important to see that all the modes $Q_{k_R}(t)$
on the infrared
(IR) sector are real. If we write these modes as a complex function with
components $u_{k_R}(t)$ and $v_{k_R}(t)$: $Q_{k_R}(t) = u_{k_R}(t)
+ i v_{k_R}(t)$,
the condition for the modes to be real is
\begin{equation}\label{alpha}
\alpha_{k_R}(t) = \left| \frac{v_{k_R}(t)}{u_{k_R}(t)}\right|
\ll 1.
\end{equation}
Hence, the condition for the field $A_{\mu}$ to be classical
on the IR sector during inflation, becomes
\begin{equation}
\frac{1}{M(t)} \sum^{k_r \simeq \vartheta k_H(t)}_{k_R=0}
\alpha_{k_R}(t) \ll 1,
\end{equation}
where $M(t)$ is the time-dependent number of degrees of freedom (which
increases with time during inflation) in the IR
($k_R \ll k_H$) sector.
The coarse-graining field
\begin{equation}
\left.{\cal A}_{\mu}\right|_{IR} = \frac{1}{(2\pi)^{3/2}} {\Large\int}
d^3k_R \Theta(\epsilon k_H - k_R) \sum^{3}_{\gamma=0}
\epsilon^{\mu}_{(\gamma)} \left[
a^{\gamma}_{k_R k_{\psi_0}} e^{i \vec k_R.\vec R}
Q_{k_R}(t) + c.c.\right],
\end{equation}
(here $\Theta$ denotes the Heaviside function),
takes into account only the modes with $10^3 \ll \vartheta^{-1}
< k_H/k_R$ that can be considered as classical because
\begin{equation}
\left[ \left.{\cal A}_{\mu}\right|_{IR} ,
\left.\dot {\cal A}_{\mu}\right|_{IR}\right]
\simeq 0,
\end{equation}
which in turn implies that the fluctuations of $\left.A_{\mu}\right|_{IR}$
can be treated as classical
in the electromagnetic field as well.

\subsection{Electromagnetic fields during inflation}

Previous results allow us to calculate the effective 4D super Hubble squared
fluctuations of the electromagnetic field
$<0|A^{\mu}A_{\mu}|0>=\left<A^{2}\right>$, which are given by
\begin{equation}\label{eq34}
\left.\left<A^{2}\right>\right|_{IR}
=\frac{2e^{-3H_{0}t}}{\pi ^2}\int _{0}^{\vartheta
k_{H}}\frac{dk_{R}}{k_{R}}k_{R}^{3}Q_{k_{R}}Q_{k_{R}}^{*},
\end{equation}
where $\vartheta =\frac{k_{max}^{(IR)}}{k_p}\ll 1$ is a dimensionless parameter.
Here, $k_{max}^{(IR)}=k_{H}(t_{i})=H_{0}e^{H_{0}t_i}$
is the wavenumber related to the Hubble radius at $t_{i}$
(the time when the horizon enters)
and $k_{p}$ is the Planckian wavenumber.
In fact we choose $k_p$ as a cut-off scale of all the spectrum.

To obtain
$\left.\left<A^{2}\right>\right|_{IR}$,
we must consider the small argument
limit for ${\cal H}_{\nu}^{(2)}[y]\simeq \frac{\left(\frac{y}{2}\right)^{\nu}}{\Gamma(1+\nu)}-\frac{i}{\pi}
\Gamma (\nu)\left(\frac{y}{2}\right)^{-\nu}$.
From the condition (\ref{alpha})
we obtain that each $k_R$-mode becomes classical for times
\begin{equation}\label{condicion}
t \gg \frac{1}{2\nu H_0} \  {\rm ln} \left[\frac{\Gamma(\nu)}{\Gamma(1+\nu)
2^{2\nu}} \left(\frac{k_R}{H_0}\right)^{2\nu}\right],
\end{equation}
which for a de Sitter expansion takes the form $\frac{k_R}{H_0} \ll e^{N_e}$
and $N_e$ is the number of e-folds at the end of inflation.
In order for inflation to solve the horizon/flatness problem,  $N_e \ge
60$ is required.
Note that in this case when $k_{\psi _0}^{2}\simeq
2H_{0}^{2}$, the constant
parameter $|\alpha |/H^2_0 \ll 1$ and thus $\nu \simeq 3/2$. Hence we
can use ${\cal H}_{\nu}^{(2)}[y]\simeq -\frac{i}{\pi}\Gamma (\nu)
\left(\frac{y}{2}\right)^{-\nu}$  on the IR
sector to obtain
\begin{equation}\label{eq35}
\left.\left<A^{2}\right>\right|_{IR} \simeq
\frac{2^{2\nu -1}}{\pi ^3}H_{0}^{2\nu -1}\Gamma ^{2}
(\nu)e^{-(3-2\nu)H_{0}t}\int _{0}^{\vartheta
k_{H}}\frac{dk_{R}}{k_R}k_{R}^{3-2\nu}.
\end{equation}
Note that when $\alpha=0$, $\nu =3/2$  and thus the
spectrum ${\cal P}(k_{R})\sim k_{R}^{3-2\nu}$ is scale invariant.
Performing
the remaining integration, (\ref{eq35}) becomes
\begin{equation}\label{eq36}
\left.\left<A^{2}\right>\right|_{IR} \simeq
\frac{2^{2\nu -1}}{\pi ^3}\frac{\Gamma ^{2}(\nu)}{(3-2\nu)}H_{0}^{2}
\vartheta^{3-2\nu},
\end{equation}
which is similar to the corresponding $\left.\left<\varphi^2\right>
\right|_{IR}$. We must note that $\left<A^{2}\right>$ has
constant value in the infrared sector, which means that the amplitude
of the corresponding photons is constant. This result can be interpreted as
a classical
large-scale electromagnetic potential generated when a de
Sitter inflationary process ends, which is responsible
for a large-scale seed mag\-ne\-tic field.

\section{Induced seed magnetic field.}

In this section we estimate the seed magnetic field induced from
the electromagnetic potential whose
dynamics was studied in the previous section.
For this purpose we consider the 3D spatial components
of $\vec A=A^i \hat e_i$ ($\hat e_i$ are the 3D spatial basis vectors),
which in view of (\ref{eq26}) and (\ref{eq28}) satisfy
\begin{equation}\label{smf1}
\ddot{A}^{i}+3H_{0}\dot{A}^{i}
-e^{-2H_{0}t}\nabla _{R}^{2}A^{i}-\alpha A^{i}=0.
\end{equation}
We consider the physical components
of $\vec{A}$ and $\vec{B}$ measured in a comoving frame. Hence, the
orthonormal basis components associated with the observers in this
frame are given by
\begin{equation}\label{smf3}
e_{(t)}=\frac{\partial}{\partial t},\qquad e_{(\bar{R})}=
e^{-H_{0}t}\frac{\partial}{\partial \bar{R}},\qquad e_{(\theta)}
=e^{-H_{0}t}\frac{1}{\bar{R}}\frac{\partial}{\partial
\theta},\qquad e_{(\phi)}= e^{-H_{0}t}\frac{1}{\bar{R} \  {\rm sin}\theta}
\frac{\partial}{\partial \phi},
\end{equation}
where the spatial part of (\ref{eq25}) has been rewritten as $dR^{2}=d\bar{R}^{2}+\bar{R}^{2}
(d\theta ^{2}+{\rm sin}^{2}\theta d\phi ^{2})$.

On the other hand, we know that from (\ref{eq9}) through
the transformations (\ref{eq23}) we
can obtain the standard Maxwell's equations with sources, where
such sources have a geometrical origin.
The Maxwell's equations without sources can be obtained from
these (see \cite{Jackson}). Therefore, using
$\vec\nabla _{R}\cdot \vec{B}_{com}=0$ and
$\vec B_{com} = \vec{\nabla}_R \times \vec{A}_{com}$,
equation (\ref{smf1}) becomes
\begin{equation}\label{smf4}
\ddot{B}^{i}_{com}+H_{0}\dot{B}^{i}_{com}-e^{-2H_{0}t}
\nabla _{R}^{2}B^{i}_{com}
-(\alpha +2H_{0}^{2})B^{i}_{com}=0.
\end{equation}
This expression describes the dynamics of the comoving components of the
seed magnetic
field. As in the case of $A^{\mu}$, we can express these components
 as a Fourier expansion
\begin{equation}\label{smf5}
B^{i}_{com}(t,\vec{R})=\frac{e^{-\frac{1}{2}H_{0}t}}{(2\pi)^{3/2}}
\int d^{3}k_{R}\sum _{l=1}^{3}\epsilon _{(l)}^{i}(k_{R})
\left[b_{k_{R}}^{(l)}e^{i\vec{k}_{R}
\cdot\vec{R}}G_{k_{R}}(t)
+b_{k_{R}}^{(l)\,\dagger}e^{-i\vec{k}_{R}\cdot\vec{R
}}G_{k_{R}}^{*}(t)\right],
\end{equation}
where $b_{k_{R}}^{(l)\,\dagger}$ and $b_{k_{R}}^{(l)}$ are the creation and
annihilation operators and $\epsilon _{(l)}^{i}(k_{R})$ are
the 3-polarisation vectors which satisfy $\epsilon _{(i)}\cdot \epsilon _{(j)}=g_{ij}$.
Therefore, the equation of motion for
$G_{k_{R}}(t)$ obtained from (\ref{smf4}), acquires the form
\begin{equation}\label{smf7}
\ddot{G}_{k_{R}}+\left[k_{R}^{2}e^{-2H_{0}t}
-\left(\frac{9}{4}H_{0}^{2}+\alpha\right)\right]G_{k_{R}}=0,
\end{equation}
and has for solution
\begin{equation}\label{smf7'}
G_{k_{R}}(t)=L_{1}{\cal H}_{\nu}^{(1)}[w(t)]+L_{2}{\cal H}_{\nu}^{(2)}[w(t)],
\end{equation}
where $\nu =\frac{1}{2H_0}\sqrt{9H_{0}^{2}+4\alpha}$,
$w(t)=\frac{k_{R}}{H_0}e^{-H_{0}t}$ and $L_1$, $L_2$ are integration
constants.
The corresponding normalization condition for those seed magnetic modes is
\begin{equation}\label{smf8}
G_{k_{R}}\dot{G}_{k_{R}}^{*}-\dot{G}_{k_{R}}G_{k_{R}}^{*}=i/a^2_0,
\end{equation}
being $a_0 = H^{-1}_0$ the scale factor of the universe when
inflation begins.

The normalized solution
of (\ref{smf7}) is
\begin{equation}\label{smf9}
G_{k_{R}}(t)=i\sqrt{\frac{\pi H_0}{4}}{\cal H}_{\nu}^{(2)}[w(t)].
\end{equation}
Now,  the super Hubble squared $B$
fluctuations of the
seed magnetic field $\left<B^{2}_{com}\right>$ in the Feynman gauge are given by
\begin{equation}\label{smf11}
\left.\left<B^{2}_{com}\right>\right|_{IR}
=\frac{3e^{-H_{0}t}}{2\pi^{2}}\int _{0}^{\vartheta k_{H}}
\frac{dk_{R}}{k_R}\, k_{R}^{3}G_{k_{R}}(t)G_{k_{R}}^{*}(t)=
\int _{0}^{\vartheta k_{H}} \frac{dk_{R}}{k_R} \  {\cal P}(k_R),
\end{equation}
where $\vartheta=\frac{k_{max}^{(IR)}}{k_{p}}\ll 1$ is a
dimensionless parameter.
We are choosing $k_{p}$ as a cut-off scale of the whole spectrum.
The power spectrum ${\cal P}(k_R)$ on
cosmological scales is
\begin{equation}
{\cal P}(k_R) = \frac{3}{8\pi^{3}}2^{2\nu}\Gamma ^{2}(\nu)H_{0}^{1+2\nu}e^{-(1-2\nu)H_{0}t}k_{R}^{3-2\nu}.
\end{equation}
Considering the case
$k_{\psi _{0}}^{2} \simeq 2H_{0}^{2}$, we see that
$|\alpha|/H^2_0 \ll 1$, and thus $\nu \lesssim 3/2$, because $\alpha=
k^2_{\psi_0} - 2 H^2_0$. This case is of physical interest since it corresponds to a nearly scale-invariant
power spectrum for $\left.\left<B^{2}_{com}\right>\right|_{IR}$.
Therefore, on the infrared IR sector, we obtain
\begin{equation}\label{smf13}
\left.\left<B^{2}_{com}\right>\right|_{IR}\simeq
\frac{3\Gamma ^{2}(\nu)}{8\pi ^3}\frac{2^{2\nu}H_{0}^{4}}{(3-2\nu)}
\vartheta ^{3-2\nu}e^{2H_{0}t}.
\end{equation}
It is remarkable in this result that $\left.\left<B^{2}_{com}
\right>\right|_{IR}$ is a growing function of time during inflation.
We  notice that the typical
infrared divergence appears when $k^2_{\psi_0} =2H^2_0$
as in the case of
the scalar field inflaton analysis for a de Sitter expansion, where the spectrum is exactly scale invariant.

On the other hand, the physical magnetic field $B_{phys}$ is related
with the comoving one as
\begin{displaymath}
B_{phys} \sim a^{-2} \  B_{com}.
\end{displaymath}
After inflation, $B_{phys}$ decreases as $a^{-2}$.
Hence, we could make an estimation for the actual strength of
the cosmological magnetic field $B^{(a)}_{phys}$
\begin{displaymath}
\left.\left<\left(B^{(a)}_{phys}\right)^2\right>^{1/2}\right|_{IR}
\simeq \left(\frac{a(t=t_0)}{a(t=t_i)}\right)^4
\left.\left<B^2_{com}(t=t_i)\right>^{1/2}\right|_{IR},
\end{displaymath}
where $B_{com}(t=t_i)$ denotes the comoving magnetic field
at the end of inflation.

\begin{figure*}

\includegraphics[totalheight=8.5cm,angle=270]{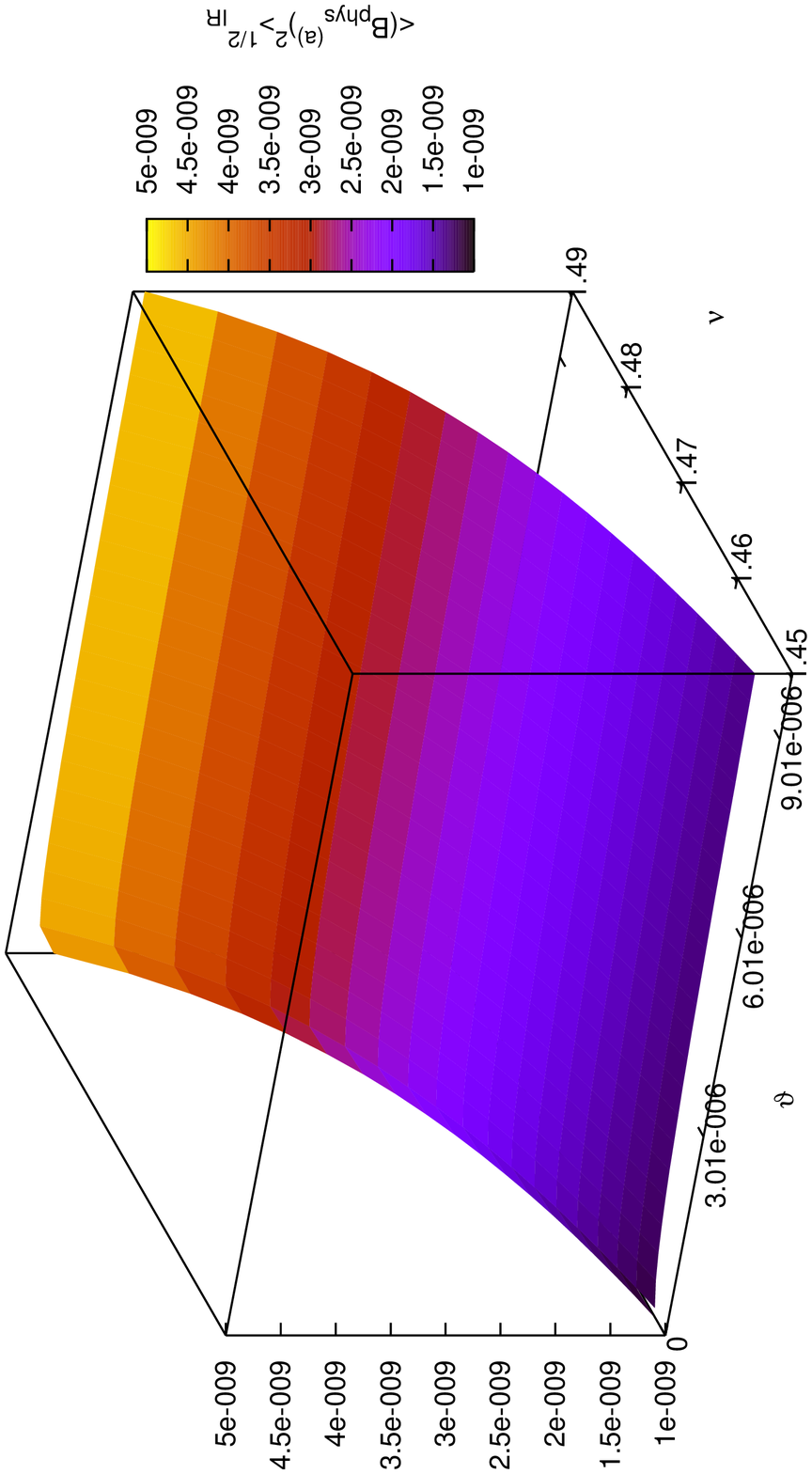}\caption{\label{f1}
$\left.\left<\left(B^{(a)}_{phys}\right)^2\right>^{1/2}\right|_{IR}$
(in Gauss), with respect to $\nu$ and $\vartheta$. Notice that
$\nu$ is related to the spectral index $n_s$ by the expression:
$n_s=4-2\nu $, so that values used in the graphic for $\nu =
(1.45, 1.49)$ correspond respectively to $n_s=(1.1, 1.02)$. Values
considered for $\vartheta$ correspond to actual scales from
$3\times 10^{3}$ to $3\times 10^{6}$ {\rm Mpc}.}
\end{figure*}

In the figure (\ref{f1}) we have plotted
$\left.\left<\left(B^{(a)}_{phys}\right)^2\right>^{1/2}\right|_{IR}$
(in Gauss),
with respect to $\nu$ and $\vartheta$. Notice that $\nu$ is related
to the spectral index $n_s$ by the expression: $n_s=4-2\nu $.
%
Furthermore, we have used
$H_0 = 0.5 \times 10^{-9} \  M_p$
taking $N_e=63$ and $\vartheta$ on the range
$10^{-5}$ to $10^{-8}$ (which corresponds to actual scales
that run from $3 \times 10^3$ to $3 \times 10^6$ {\rm Mpc}. To estimate
the scale factor evolution of
$\left.\left<\left(B^{(a)}_{phys}\right)^2\right>^{1/2}\right|_{IR}$,
we used
\begin{displaymath}
\left(\frac{a(t=t_0)}{a(t=t_i)}\right)^4 \simeq 10^{-136},
\end{displaymath}
which accounts for the actual size of the observable horizon
($\sim 10^{28} \  cm$) and the size of the horizon at the
end of inflation $(\sim 3.6 \times 10^{-6} \  cm)$.

\section{Final Comments}

In this letter we have developed a novel formalism of inflation
which takes into account gravitoelectromagnetic effects from a 5D
vacuum state, where the fifth (spatial like) coordinate is considered
as noncompact.
The reader can see a different approach in the framework of STM
theory, for instance, in \cite{OW}.
In our case, to define the 5D vacuum on the Riemann flat ($R^A_{BCD}=0$) metric
(\ref{eq1}), we introduce the density Lagrangian (\ref{eq3}),
which is purely kinetic, for a tensorial operator
\begin{displaymath}
Q_{BC} = F_{BC} + g_{BC} \left(A^{D}_{;D}\right),
\end{displaymath}
(such that $F_{BC}=A_{C;B} - A_{B;C}$ is antisymmetric and $g_{AB}$ is
symmetric)
where the vector potential has components $A_{B} = (A_{\mu},\varphi)$,
which are minimally coupled to gravity.
Working in the Feynman gauge, we obtain a 5D massless Klein-Gordon-like
equation for $A^B$, which represents the Maxwell's equations in a 5D
vacuum state (see eq. (\ref{eq6})). Using transformations (\ref{eq23})
with the foliation $\psi=H^{-1}_0$, we obtain the Maxwell's equations
on an effective 4D de Sitter background metric (\ref{eq25}), where the
sources (the last terms in (\ref{eq26}) and (\ref{eq26'})) describe
the derivatives of the corresponding potentials with respect to $A^{\mu}$ and
$\varphi$. Hence, the effective 4D dynamics of $A^{\mu}$ and the
inflaton field $\varphi$ is well described by equations (\ref{eq26})
and (\ref{eq26'}).
Finally, we have studied the evolution of the squared $B_{com}$-fluctuations
during inflation, which are classical on cosmological scales. These
fluctuations increase exponentially on cosmological scales and
at the end of this epoch its strength is of the order of $\left(
10^{127}\right)^2 \left(G\right)^2$. Later, we have
estimated the present day strength of $\left<\left(B^{(a)}_{phys}\right)^2
\right>^{1/2}$, which results of the order of $10^{-9} \  Gauss$.
This results agree with the limits imposed by the high isotropy of the CMB
photons, obtained from the COBE data\cite{mar}. However, must be
noted that our calculations are very sensitive with the number of
e-folds that one consider during inflation.

\vskip .2cm
\centerline{\bf{Acknowledgements}}
\vskip .2cm
JEMA acknowledges CONACyT and IFM of UMSNH (M\'exico)
for financial support.
and MB acknowledges CONICET
and UNMdP (Argentina) for financial support.\\


\end{document}